\documentclass[12pt]{JHEP}
\usepackage{amsmath,epsfig}
\usepackage{amssymb,amsfonts}
\usepackage{latexsym}
\usepackage{epsfig}
\def\be{\begin{equation}}
\def\ee{\end{equation}}
\def\bea{\begin{eqnarray}}
\def\eea{\end{eqnarray}}

\title{Logarithmic correction in the deformed ${\rm AdS}_5$ model
to produce the heavy quark potential and QCD beta function}
\author{Song He$^{1}$,
Mei Huang$^{1,2}$, Qi-Shu Yan$^{3,4}$ \\
$^{1}$ {Institute of High Energy Physics, Chinese Academy of
Sciences, Beijing, China} \\
$^{2}$ {Theoretical Physics Center for Science Facilities, Chinese
Academy of Sciences, Beijing, China} \\
$^{3}$ {Department of Physics, University of Toronto, Toronto,
Canada}
\\
$^{4}$ {College of Physical Sciences, Graduate University of Chinese
Academy of Sciences, Beijing, China} }

\abstract{We stude the \textit{holographic} QCD model which contains
a quadratic term $ -\sigma z^2$ and a logarithmic term
$-c_0\log[(z_{IR}-z)/z_{IR}]$ with an explicit infrared cut-off
$z_{IR}$ in the deformed ${\rm AdS}_5$ warp factor. We investigate
the heavy quark potential for three cases, i.e, with only quadratic
correction, with both quadratic and logarithmic corrections and with
only logarithmic correction. We solve the dilaton field and dilation
potential from the Einstein equation, and investigate the
corresponding beta function in the G{\"u}rsoy -Kiritsis-Nitti (GKN)
framework. Our studies show that in the case with only quadratic
correction, a negative $\sigma$ or the Andreev-Zakharov model is
favored to fit the heavy quark potential and to produce the QCD
beta-function at 2-loop level, however, the dilaton potential is
unbounded in infrared regime. One interesting observing for the case
of positive $\sigma$, or the soft-wall ${\rm AdS}_5$ model is that
the corresponding beta-function exists an infrared fixed point. In
the case with only logarithmic correction, the heavy quark Cornell
potential can be fitted very well, the corresponding beta-function
agrees with the QCD beta-function at 2-loop level reasonably well,
and the dilaton potential is bounded from below in infrared. At the
end, we propose a more compact model which has only logarithmic
correction in the deformed warp factor and has less free
parameters.}


\begin{document}

\def\g{\gamma}
\def\go{\g_{00}}
\def\gi{\g_{ii}}

\maketitle 

\section{Introduction}

Quantum Chromodynamics (QCD) has been accepted as the basic theory
of describing strong interaction for more than 30 years. However, it
is still a challenge to solve QCD in non-perturbative region where
gauge interaction is strong. Recently, the conjecture of the
gravity/gauge duality \cite{dual} has revived the hope of
understanding QCD in strongly coupled region using string theory.
The AdS/CFT duality has been widely used to discuss the meson
spectra \cite{EKSS2005,TB:05,DaRold2005} and dense and hot quark
matter
\cite{Kovtun:2004de,etas-adscft,bulk-adscft,polyakov-adscft,densematter-adscft}.
The string description of realistic QCD has not been successfully
formulated yet. Many efforts are invested in searching for such a
realistic description by using the "top-down" approach,
\textit{i.e.} by deriving holographic QCD from string theory
\cite{D3-D7,D4-D6,SS,Dp-Dq}, as well as by using the
"bottom-up" approach, \textit{i.e.} by examining possible \textit{holographic%
} QCD models from experimental data and lattice results.

In the "bottom-up" approach, the most economic way might be to
search for a deformed ${\rm AdS}_5$ metric
\cite{KKSS2006,Andreev:2006ct,Andreev:2006vy,Shock-2006,Ghoroku-Tachibana,
Gursoy,Zeng:2008sx,Pirner:2009gr}, which can describe the known
experimental data and lattice results of QCD, e.g. hadron spectra
and the heavy quark potential. The simplest holographic QCD model is
the hard-wall ${\rm AdS}_5$ model
\cite{Hardwall-Polchinski,EKSS2005}, which can describe the lightest
meson spectra in $80-90\%$ agreement with the experimental data.
However, the hard-wall model cannot produce the Regge behavior for
higher excitations. It is regarded that the Regge behavior is
related to the linear confinement. It has been suggested in Ref.
\cite{KKSS2006} that a negative quadratic dilaton term $-z^2$ in the
action is needed to produce the right linear Regge behavior of
$\rho$ mesons or the linear confinement.

The most direct physical quantity related to the confinement is the
heavy-quark potential. The lattice result which is consistent with
the so called Cornell potential \cite{Cornell} has the form of
\begin{equation}
V_{Q{\bar Q}}(R)=-\frac{\kappa}{R}+\sigma_{str}R+V_0.
\label{quarkpotential}
\end{equation}
Where $\kappa\approx 0.48$, $\sigma_{str}\approx 0.183 {\rm
GeV}^{2}$ and $V_0=-0.25 {\rm GeV}$, the first two parameters can be
interpreted as $\frac{4\alpha_s}{3}$ and QCD "string" tension,
respectively.

In order to produce linear behavior of heavy flavor potential,
Andreev and Zakharov in Ref.\cite{Andreev:2006ct} suggested a
positive quadratic term modification \cite{Andreev:2006vy} in the
deformed warp factor of the metric, which is different from the
soft-wall model in \cite{KKSS2006}. Andreev-Zakharov model has been
further studied in many other articles
\cite{Jugeau:2008ds,Zuo:2009dz,deTeramond:2009xk,White:2007tu}. In
Ref. \cite{White:2007tu}, the authors found that the heavy quark
potential from the positive quadratic model is closer to the Cornell
potential than that from the backreaction model \cite {Shock-2006},
which contains higher order corrections.

It is clearly seen from the Cornell potential that the Coulomb
potential dominates in the ultraviolet (UV) region and the linear
potential dominates in the infrared (IR) region. It motivates people
to take into account the QCD running coupling effect into the
modified metric \cite{Gursoy,Zeng:2008sx,Pirner:2009gr}. In
Ref.\cite{Pirner:2009gr}, Pirner and Galow have proposed a deformed
metric which resembles the QCD running coupling, and the
Pirner-Galow metric can fit the Cornell potential reasonably well.
However, as shown in Ref. \cite{Galow:2009kw} the corresponding
dilaton potential solved from the Einstein equation is unstable, and
the corresponding beta function does not agree with the QCD beta
function.

The motivation of this paper is to search for a deformed ${\rm
AdS}_5$ metric, which can describe the heavy quark potential as well
as the QCD $\beta$ function and at the same time can have a stable
dilaton potential from the gravity side. In \cite{Dp-Dq}, we have
proposed the soft-wall Dp-Dq model, which contains a quadratic
correction and a logarithmic correction $c_0\log z$. The logarithmic
dilaton correction is derived from the top-down method, which is
general for $Dp-Dq$ system except $p=3$. As pointed in
\cite{Gursoy}, that the logarithmic term $c_0\log z$ itself cannot
produce confinement, while a logarithmic correction with an infrared
cut-off in the form of $c_0\log (z_{IR}-z)$ can have confinement at
IR. Therefore, we propose a \textit{holographic} QCD model, which
contains a quadratic term $-\sigma z^2$ and a logarithmic term
$-c_0\log[(z_{IR}-z)/z_{IR}]$ with an explicit infrared cut-off
$z_{IR}$ in the deformed ${\rm AdS}_5$ warp factor, where we assume
$\sigma, c_0$ can be either positive or negative. This model is
found to have the same metric structure of the Pirner-Galow's model
\cite{Pirner:2009gr} in UV and IR region respectively. We
investigate the heavy quark potential in the proposed model for
three cases, i.e, with only quadratic correction, with both
quadratic and logarithmic corrections and with only logarithmic
correction. We solve the dilaton field and dilation potential from
the Einstein equation, and investigate the corresponding beta
function in the G{\"u}rsoy -Kiritsis-Nitti (GKN) \cite{Gursoy}
framework . Our studies show that in the case with only quadratic
correction, the produced heavy quark potential has both Coulomb part
and linear part and fits the Cornell potential qualitatively well.
In the case with only logarithmic correction, the heavy quark
Cornell potential can be perfectly fitted and the corresponding
beta-function agrees with the QCD beta-function reasonably well. At
the end, we propose a more compact model which has only logarithmic
correction in the deformed warp factor and has less free parameters.

The paper is organized as follows. In section II, we derive the
general formulae for heavy quark potential in the framework of
AdS/CFT, and introduce the GKN framework to construct the gravity
dual of the 5D holographic model, and calculate the $\beta$
function. In section III, we construct our \textit{holographic} QCD
model, which contains a quadratic term $-\sigma z^2$ and a
logarithmic term $-c_0\log[(z_{IR}- z)/z_{IR}]$ with an explicit
infrared cut-off $z_{IR}$ in the deformed ${\rm AdS}_5$ warp factor.
We fit the heavy quark potential in this model for three cases, i.e,
with only quadratic correction, with both quadratic and logarithmic
corrections and with only logarithmic correction. We solve the
dilaton field and dilation potential from the Einstein equation, and
investigate the corresponding beta function. In section IV, we
propose a more compact model with only logarithmic correction and
with less parameters. The summary and discussion is given in section
V.

\section{The formalism}

The ${\rm AdS}_5$ metric in the Euclidean space takes the form of
\begin{eqnarray}
ds^2=G_{\mu\nu}^s dX^\mu dX^\nu &=& \frac{L^2}{z^2}\left(
dt^2+d\vec{x}^2+dz^2\right), \label{metirc-ads5}
\end{eqnarray}
where $G_{\mu\nu}^s$ indicates the metric in the string frame, and
$L$ is the radius of ${\rm AdS}_5$. To search for the possible
\textit{holographic} QCD models, the most economic way of breaking
conformal invariance is to add a deformed warp factor $h(z)$ in the
metric background, and the general metric $\mathcal {A}(z)$ in
Euclidean space has the following form:
\begin{eqnarray}
ds^2=G_{\mu\nu}^s dX^\mu dX^\nu &=& \frac{h(z)L^2}{z^2}\left(
dt^2+d\vec{x}^2+dz^2\right) \label{h-general} \\
&=& e^{2\mathcal {A}(z)}\left( dt^2+d\vec{x}^2+dz^2\right).
\label{metric-general}
\end{eqnarray}
In this context, we introduce the warp factor into the pure ${\rm
AdS}_5$ to break the conformal symmetry to find the QCD-like gauge
theory. We will derive the general formula for heavy quark potential
and work out the dilaton potential
 which will be useful later.

\subsection{The heavy quark potential from $AdS/CFT$}
\label{sec-HQ-Wilsonloop}

To keep the paper self-contained, we follow the standard procedure
\cite{Wilsonloop-adscft} to derive the static heavy quark potential
$V_{Q{\bar Q}}(R)$ under the general metric background of
(\ref{metric-general}). In $SU(N)$ gauge theory, the interaction
potential for infinity massive heavy quark antiquark is calculated
from the Wilson loop
\begin{equation}
W[C]=\frac{1}{N} Tr P \exp[i \oint_{C} A_\mu dx^\mu],
\label{Wilson-loop-formula}
\end{equation}
where $A_{\mu}$ is the gauge field, the
trace is over the fundamental representation, $P$ stands for path ordering.
$C$ denotes a closed loop in spacetime, which is a rectangle with
one direction along the time direction of length $T$ and the other
space direction of length $R$. The Wilson loop describes the
creation of a $Q{\bar Q}$ pair with distance $R$ at some time
$t_0=0$ and the annihilation of this pair at time $t=T$. For
$T\to\infty$, the expectation value of the Wilson loop behaves as
$\langle W(C)\rangle\propto e^{-T V_{Q\bar Q}}$.

According to the \textit{holographic} dictionary, the expectation
value of the Wilson loop in four dimensions should be equal to the
string partition function on the modified ${\rm AdS}_5$ space, with
the string world sheet ending on the contour $C$ at the boundary of
${\rm AdS}_5$
\begin{equation}
\langle W^{4d}[C]\rangle=Z_{string}^{5d}[C]\simeq e^{-S_{NG}[C]} \,\
,
\end{equation}
where $S_{NG}$ is the classical world sheet Nambu-Goto action
\begin{equation}
S_{NG}=\frac{1}{2\pi\sigma_s}\int d^2 \eta \sqrt{{\rm Det} \chi_{a
b}},
\end{equation}
with $\sigma_s$ the string tension which has dimension of ${\rm
GeV}^{-2}$, and $\chi_{ab}$ is the induced worldsheet metric with
$a,b$ the indices in the ($\eta^0=t,\eta^1=x)$ coordinates on the
worldsheet. Under the background (\ref{metric-general}), we can
obtain the equation of motion:
\begin{equation}
 \frac{e^{2\mathcal
{A}(z)}}{\sqrt{1+(z')^2}}=e^{2\mathcal {A}(z_0)},
\end{equation}
where $z_0$ is the maximal value of z. Following the standard
procedure, one can derive the interquark distance $R$ as a function
of $z$
\begin{eqnarray}
R(z) &=&2 z \int_0^1 d\nu \frac{e^{2\mathcal {A}(z)}}{e^{2\mathcal
{A}(\nu z)}}\frac{1}{\sqrt{1- \left (\frac{e^{2\mathcal
{A}(z)}}{e^{2\mathcal {A}(\nu z)}}\right )^2}}. \label{distance1}
\end{eqnarray}
The heavy quark potential can be worked out from the Nambu-Goto
string action:
\begin{eqnarray} V_{Q\bar
Q}(z)&=&\frac{1}{\pi\sigma_s}\int_0^1 d\nu e^{2\mathcal {A}(\nu
z)}z\frac{1}{\sqrt{1-\left ( \frac{e^{2\mathcal
{A}(z)}}{e^{2\mathcal {A}(\nu z)}}\right )^2}}. \label{VQQ-general}
\end{eqnarray}
It is noticed that the integral in Eq.(\ref{VQQ-general}) in
principle include some poles, which induces $ V_{Q\bar
Q}(z)\rightarrow \infty$. The infinite energy should be extracted
through certain regularization procedure. The divergence of
$V_{Q\bar Q}(z)$ is related to the vacuum energy for two static
quarks. Generally speaking, the vacuum energy of two static quarks
will be different in various background. In our latter calculations,
we will use the regularized $V_{Q\bar Q}^{ren.}$, for example Eqs.
(\ref{Vregular1}) and (\ref{Vregular2}), where the vacuum energy has
been subtracted.

\subsection{The GKN framework of the gravity dual theory}
\label{gravity}

Motivated from finding the appropriate description of heavy quark
potential from gravity theory side, we expect to work out a general
classical gravity background. If the deformed ${\rm AdS}_5$ metric
can describe QCD phenomenology, it is natural to ask whether it is
possible to find its dual theory from gravity side. The G{\"u}rsoy
-Kiritsis-Nitti (GKN) \cite{Gursoy} framework offers a systematical
procedure to construct the gravity dual theory for a 5D holographic
QCD model defined in Eq.(\ref{metric-general}).

In this paper, we follow the notation in Ref.\cite{Galow:2009kw} to
introduce the GKN framework. According to the GKN's framework, the
noncritical string background dual to the QCD-like gauge theories
can be described by the following action in the Einstein frame:
\begin{equation}
S_{5D-Gravity}=\frac{1}{2\kappa_5^2}\int d^5x \sqrt{-G^E} \left (R
-\frac{4}{3}\partial_{\mu}\phi\partial^{\mu}\phi-V_B(\phi)\right)\,.
\label{5D_action}
\end{equation}
Where $R$ is the Ricci scalar and has the dimension
$[R]=\frac{1}{\mbox{length}^2}$, $\kappa_5^2$ has dimension
$[\kappa_5^2]=\mbox{length}^3$, $\phi$ is the dilaton field and is
dimensioness, $V_B(\phi)$ the dilaton potential and has the
dimension of $[V_B(\phi)]=\frac{1}{\mbox{length}^{2}}$. The metric
in the Einstein frame is denoted by $G_{\mu\nu}^E$ which is related
by the metric in the string frame $G_{\mu\nu}^s$ by the following
relation:
\begin{equation}
G_{\mu\nu}^E(X)=e^{-\frac{4}{3}\phi}G_{\mu\nu}^s(X)\,.
\label{Einstein_metric}
\end{equation}

In this subsection, the space-time metric has Minkowski signature
with the sign convention $(-,+,+,+,+)$.
\begin{equation}
ds_E^2=e^{2A(z)}(-dt^2+d\vec{x}^2+dz^2),
\label{Einstein_metric_5D_1}
\end{equation}
here the warp factor $A(z)$ is in the Einstein frame, which is
related to $\mathcal{A}(z)$ in the string frame by
\begin{equation}
e^{2A(z)}=e^{-\frac{4}{3}\phi}
\frac{h(z)L^2}{z^2}=e^{-\frac{4}{3}\phi} e^{2\mathcal{A}(z)}.
\end{equation}

The two independent Einstein's equations take the following form,
\begin{eqnarray}
3((A'(z))^2+A''(z))=-\frac{2}{3}(\phi')^2-\frac{1}{2}e^{2A(z)}V_B(\phi),
\label{phi_prime}\\
6(A'(z))^2=\frac{2}{3}(\phi')^2-\frac{1}{2}e^{2A(z)}V_B(\phi).
\label{V_phi}
\end{eqnarray}
Adding these equations one can obtain a formal expression for the
dilaton potential:
\begin{equation}
V_B(\phi(z))=-e^{-2A(z)} (9(A'(z))^2+3A''(z)). \label{DilatonVphi}
\end{equation}
Subtracting Eq.~(\ref{phi_prime}) from Eq.~(\ref{V_phi}), one can
find an important relation between the dilaton and the metric
profile:
\begin{equation}
(\phi')^2=\frac{9}{4}((A'(z))^2-A''(z)). \label{Ephiz}
\end{equation}
It is noticed that Eq.~(\ref{Ephiz}) depends on the profile $A(z)$,
which is a function of the deformed warp factor $h(z)$ and the
dilaton field $\phi(z)$. The resulting second order differential
equation for $\phi(z)$ needs two boundary conditions, which we will
obtain from the QCD running coupling constant once the bulk
coordinate $z$ is connected with the energy scale $E(z)$.

\subsection{The running coupling and the beta function}
\label{betafunction}

In the GKN framework, the scalar filed or dilaton field $\phi$
encodes the running of the Yang-Mills gauge theory's coupling
$\alpha$. For convenience, the renormalized dilaton field $\phi$ has
been defined as
\begin{equation}
  \alpha= e^{\phi} \,.
\end{equation}

The warping of the bulk space relates the bulk coordinate $z$ to the
energy scale $E(z)$ via the gravational blue-shift
\cite{Galow:2009kw}.  By using the radial coordinate $r \propto
1/z$, the blue-shift is given by the dimensionless ratio
\begin{equation}
\frac{E_r}{E_{r\to\infty}}=\sqrt{\frac{G_{tt}(r\to\infty)}{G_{tt}(r)}},
\end{equation}
where $G_{tt}$ denotes the temporal component of the metric. In the
limit $r\to\infty$, the space-time is asymptotically flat, and
$G_{tt}(r\to\infty)=-1$. Hence, the blue-shift reads
\begin{equation}
E_{r\to\infty}=E_r\sqrt{-G_{tt}(r)}\quad\text{or equivalently}\quad
E_{r\to\infty}=E_z\sqrt{-G_{tt}(z)}.
\end{equation}
A simplified expression for the energy scale in the Einstein frame
has been given in \cite{Galow:2009kw}, which has the following form:
\begin{align}
  E_{r\to\infty} & = e^{-\frac{2}{3}\phi(z)}\frac{\sqrt{h(z)}}{z}
\label{Energy_gauge_1}\\
  {} & = \alpha^{-\frac{2}{3}}\frac{\sqrt{h(z)}}{z}
\label{Energy_gauge_2}
\end{align}

If one knows the value of the coupling constant $\alpha$ at a given
energy scale $E=E_{r\to\infty}$, one can find the corresponding
value of $z$ from Eq.~(\ref{Energy_gauge_2}). Then at given value of
$z$, one can obtain $\phi(z)=\mathrm{log}(\alpha)$. In order to
solve Eq.~(\ref{Ephiz}), two boundary conditions are needed. In
\cite{Galow:2009kw}, the authors have chosen two points of running
coupling $\alpha(E)$ from PDG \cite{PDG}.

The $\beta$-function has the definition of
\begin{equation}
  \beta\,\equiv\,E\frac{d\alpha }{d E}.
\end{equation}
For a 5D holographic model, its $\beta$ function is related to the
deformed warp factor $h(z)$ by
\begin{equation}
  \beta\,\equiv\,E\frac{d\alpha }{d
E}=\frac{e^\phi d\phi}{d A}=\frac{e^{\phi(z)}\cdot \phi'(z)}{A'(z)}.
\label{calculatebeta}
\end{equation}

As we know, the QCD $\beta$-function at 2-loop level has the
following form:
\begin{equation}
  \beta(\alpha)=-b_0 \alpha^2 - b_1 \alpha^3,
\label{QCDbeta}
\end{equation}
with $b_0=\frac{1}{2\pi}(\frac{11}{3}N_c-\frac{2}{3}N_f)$, and
$b_1=\frac{1}{8\pi^2}(\frac{34}{3} N_c^2-(\frac{13}{3}
N_c-\frac{1}{N_c})N_f)$ \cite{vanRitbergen:1997va}. We choose
$N_c=3$ and $N_f=4$. In this case, $b_0=\frac{25}{6\pi}$, and
$b_1=\frac{77}{12\pi^2}$.

The yielded beta function in \cite{Galow:2009kw} does not
monotonically decrease with $\alpha$. We will show in Sec. III that
this behavior of beta function can be improved by choosing different
boundary conditions.


\section{The \textit{holographic} QCD model with quadratic and
logarithmic corrections}

In \cite{Dp-Dq}, we have derived the $Dp-Dq$ model from top-down
method, and found that for any $Dp-Dq$ system except $p=3$, there is
a general logarithmic dilaton background field.  In order to
generate the Regge behavior for the light flavor mesons, we have
proposed the soft-wall Dp-Dq model, which contains a quadratic
correction and a logarithmic correction $c_0\log z$. As pointed in
\cite{Gursoy}, that the logarithmic term $c_0\log z$ itself cannot
produce confinement, while a logarithmic correction with an infrared
cut-off in the form of $c_0\log (z_{IR}-z)$ can have confinement at
IR. Therefore, we extend our soft-wall $Dp-Dq$ model to the
following form with the deformed warp factor as
\begin{equation}
 h(z)=\exp\left( -\frac{\sigma
 z^2}{2}-c_0\ln(\frac{z_{IR}-z}{z_{IR}})\right).
 \label{metric-firstmodel}
\end{equation}
The coefficients $\sigma$ and $c_0$ can be either positive or
negative. An IR cut-off $z_{IR}$ explicitly sets in the metric,
which has the same effect as the hard-wall model \cite{EKSS2005}.
When $c_0=0$, $\sigma>0$ and $\sigma<0$ corresponds to the soft-wall
model \cite{KKSS2006} and Andreev model, respectively.

In Ref.\cite{Pirner:2009gr}, in order to mimic the QCD running
coupling behavior, Pirner and Galow proposed the deformed warp
factor
\begin{equation}
h_{PG}(z)=\frac{\log\left (\frac{1}{\epsilon} \right )}{\log\left
[\frac{1}{(\Lambda z)^2+\epsilon}\right ]}.
\label{metric-PG}
\end{equation}
This metric with asymptotically conformal symmetry in the UV and
infrared slavery in the IR region yields a good fit to the heavy
$Q\bar Q$-potential with $\Lambda=264\,\text{MeV}$ and
$\epsilon=\Lambda^2 l_s^2=0.48$. It is worthy of mentioning that the
deformed warp factor $h_{PG}(z)$ is dominated by a quadratic term $
\sigma z^2$ in the UV regime and a logarithmic term
$-\log(z_{IR}-z)$ in the IR regime, respectively. The deformed
metric in Eq.(\ref{metric-firstmodel}) when taking the parameter of
$\sigma=0.08, c_0=1, z_{IR}=2.73 {\rm GeV}^{-1}$ can mimic the
Pirner-Galow deformed metric in Eq.(\ref{metric-PG}).

Under the background (\ref{metric-firstmodel}), the derived heavy
quark potential, after subtracted the vacuum energy, has the form of
\begin{equation}\label{Vregular1}
\begin{split}
V_{Q\bar Q}^{ren.}(z)=-\frac{1}{\pi\sigma_s}\frac{L^2}{z}
+\frac{1}{\pi\sigma_s}\frac{L^2}{z}\int_0^1 d\nu \left (\frac{h(\nu
z)}{\nu^2}\vphantom{\frac{1}{\sqrt{1-\nu^4 \left( \frac{h(z)}{h(\nu
z)} \right)^2}}-\frac{1}{\nu^2}}\right. \left.
\frac{1}{\sqrt{1-\nu^4\left ( \frac{h(z)}{h(\nu z)}\right
)^2}}-\frac{1}{\nu^2}-\frac{c_0 z}{z_{IR} \nu}\right ),
\end{split}
\end{equation}
and the distance between the quark-antiquark $R$ has the form of
\begin{equation}
R(z)=2 z \int_0^1 d\nu \nu^2 \frac{h(z)}{h(\nu
z)}\frac{1}{\sqrt{1-\nu^4 \left (\frac{h(z)}{h(\nu z)}\right )^2}}.
\label{distance}
\end{equation}
In the UV limit, i.e, $z\rightarrow 0$, the heavy quark potential
has the following simplified expression:
\begin{eqnarray}
V_{Q\bar Q}^{UV}(R)&=&-\frac{0.23 L^2}{\sigma_s R}+\frac{0.17
c_0 L^2}{\sigma_s z_{\text{IR}}}\nonumber \\
& + & \frac{\left(0.22 c_0+0.24 c_0^2-0.22 \sigma
z_{\text{IR}}^2\right) L^2 R}{\sigma_s z_{\text{IR}}^2}.
\label{VQQ-UV}
\end{eqnarray}
It is noticed that the coefficient of the Coulomb part is solely
determined by the string tension $\sigma_s$, to fit the Cornell
potential in UV regime, one can get $\sigma_s=0.38 {\rm GeV}^{-2}$.
It is also noticed that even in UV limit, both the quadratic and
logarithmic corrections contribute to the linear potential. If
$c_0>0$ and $\sigma>0$, the contribution to the linear potential
from the quadratic term and logarithmic term compete with each
other.

\subsection{With only quadratic correction}

We firstly consider the case with only quadratic correction when
$c_0=0$.

\subsubsection{The heavy quark potential}

\begin{figure}[h]
\begin{center}
\epsfxsize=6.5 cm \epsfysize=6.5 cm \epsfbox{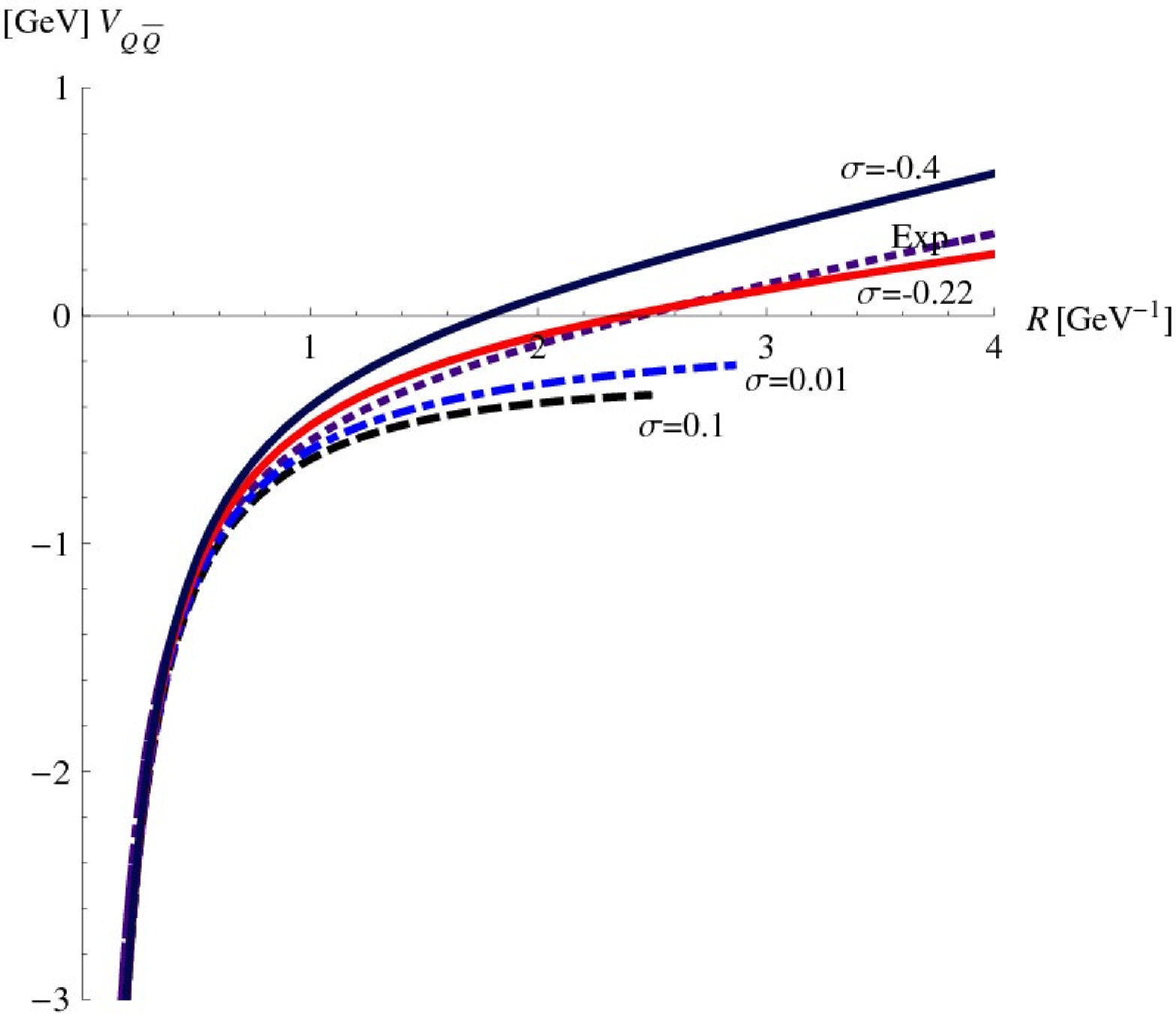}
\hspace*{0.1cm} \epsfxsize=5.5 cm \epsfysize=6.0 cm
\epsfbox{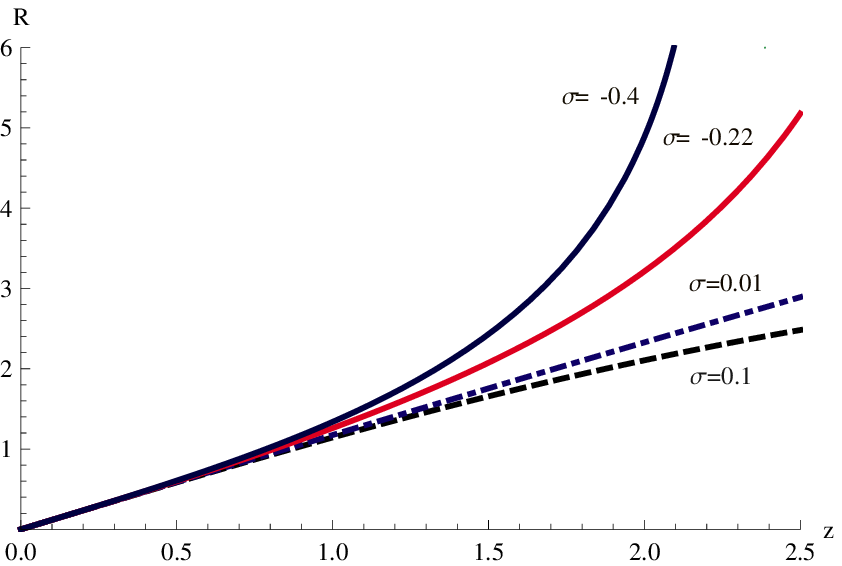} \vskip -0.05cm \hskip 0.15 cm
\textbf{( a ) } \hskip 6.5 cm \textbf{( b )} \\
\end{center}
\caption{ (a) The heavy quark potential as functions of $R$ and (b)
the distance $R$ as functions of $z$ in the case of $L=1 {\rm
GeV}^{-1}$, $\sigma_s=0.38 {\rm GeV}^{-2}$, $c_0=0$, and
$\sigma=0.1,0.01,-0.22,-0.4 {\rm GeV}^2$. }
 \label{Vqq-Rz-c0zero}
\end{figure}

 The heavy quark potential as functions of quark
anti-quark distance $R$ for different values of
$\sigma=0.1,0.01,-0.22,-0.4 {\rm GeV}^2$ is shown in Fig.
\ref{Vqq-Rz-c0zero} (a), and the corresponding relation between $R$
and $z$ is shown in Fig.\ref{Vqq-Rz-c0zero} (b). In the numerical
calculations, we have chosen the ${\rm AdS}_5$ radius $L=1{\rm
GeV}^{-1}$, and the Coulomb part is fixed by choosing the string
tension $\sigma_s=0.38$.

The behavior of $R(z)$ is quite different for $\sigma>0$ and
$\sigma<0$. In the case of $\sigma>0$, the interquark diatance $R$
firstly increases with $z$ and reach the maximum $R_m$ at certain
$z_m$, then decreases with $z>z_m$. In the case $\sigma<0$, $R$
diverges at some value of $z$ (this point is defined as $z_p$), the
heavy quark potential also diverges at $z_p$. When $c_0=0$, it can
be estimated that $z_p=\sqrt{-2/\sigma}$. Of course, in real QCD
system, the quark anti-quark cannot be separated to infinity. From
the experimental results, the linear behavior breaks around $R=1.1
{\rm fm}=5.5 {\rm GeV}^{-1}$, which is about the string breaking
scale \cite{stringbreaking}.

When $\sigma>0$, it is found that the quark anti-quark distance $R$
cannot reach the far IR regime. The larger the $\sigma$ is, the
smaller $R_m$ can be reached. The largest $R_m$ for the case of
$\sigma>0$ is around $3 {\rm GeV}^{-1}$, which is around $0.8 {\rm
fm}$. The slope for the linear potential in the middle $R$ regime is
found to be much smaller than $\sigma_{str}\approx 0.183$ in the
Cornell potential.

When $\sigma<0$, it is found that the interquark distance $R$ can go
to far IR regime. The slope of the linear potential increases with
the absolute value of $|\sigma|$. The best fit of the heavy quark
potential gives $\sigma=-0.22 {\rm GeV}^2 $. With these parameters,
the interquark distance and heavy quark potential diverges at around
$z_p=3.0 {\rm GeV}^{-1}$. However, it is noticed that in this case,
the slope of the linear potential is smaller than the experimental
value. For the case of $\sigma=-0.4$, the linear part is parallel to
the Cornell potential, however, the value of $V_{Q{\bar Q}}$ is
larger than the experimental results. Therefore, strictly speaking,
the Cornell potential is not fitted very well in the case with only
quadratic correction.

\subsubsection{The dilaton potential and the $\beta$ function for negative $\sigma$}
\label{Sec-Vphi-Beta-c0zero}

\begin{figure}
\begin{center}
\epsfxsize=6.5 cm \epsfysize=6.5 cm \epsfbox{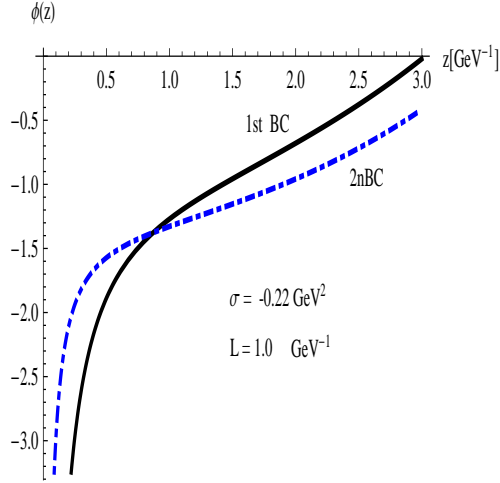}
\hspace*{0.1cm}
\end{center}
\caption[]{The dilaton field $\phi$ as a function of the the bulk
coordinate $z$ in the case of $L=1 {\rm GeV}^{-1}$, $\sigma_s=0.38
{\rm GeV}^{-2}$, $c_0=0$, and $\sigma=-0.22 {\rm GeV}^2$. The
boundary conditions are described in Eq.(\ref{BC-c0zero}). }
\label{phiz-c0zero}
\end{figure}

\begin{figure}
\begin{center}
\epsfxsize=6.5 cm \epsfysize=6.5 cm \epsfbox{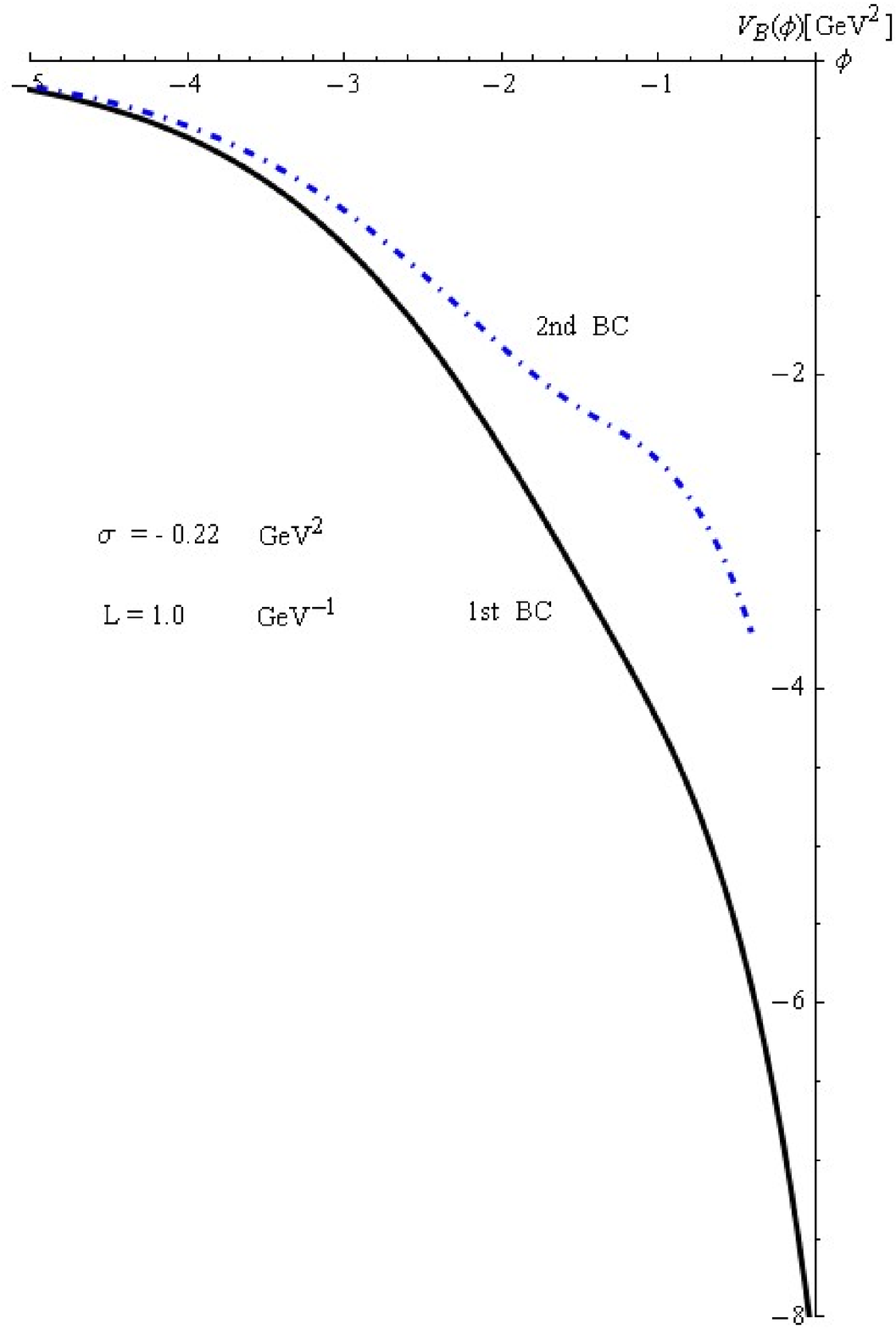}
\hspace*{0.1cm} \epsfxsize=6.5 cm \epsfysize=6.5 cm
\epsfbox{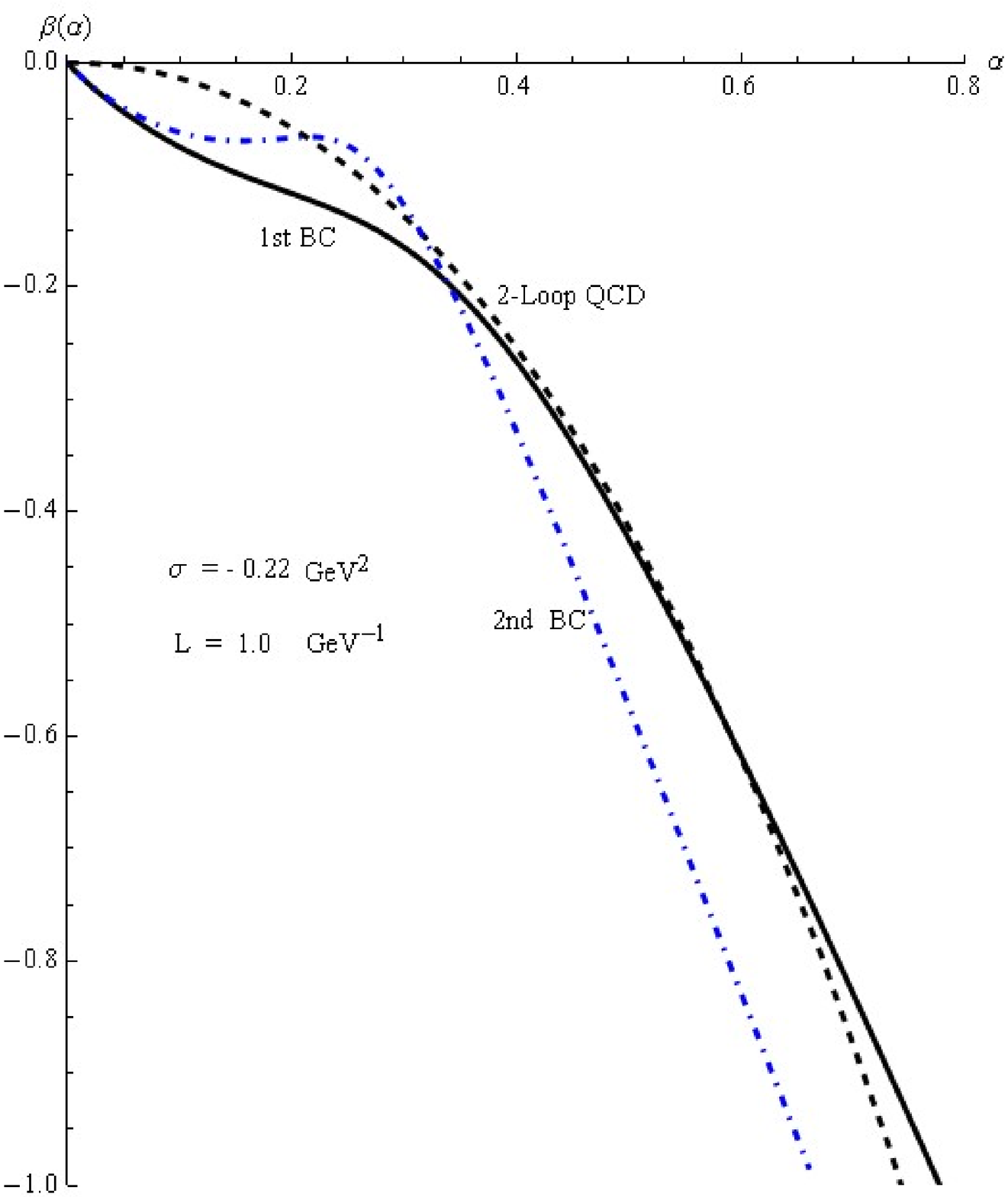} \vskip -0.05cm \hskip 0.15 cm
\textbf{( a ) } \hskip 6.5 cm \textbf{( b )} \\
\end{center}
\caption[]{ (a) The dilaton potential $V_B(\phi)$ as a function of
$\phi$ and (b) The $\beta$ function as a function of $\alpha$ in the
case of $L=1 {\rm GeV}^{-1}$, $\sigma_s=0.38 {\rm GeV}^{-2}$,
$c_0=0$, and $\sigma=-0.22 {\rm GeV}^2$. The boundary conditions are
described in Eq.(\ref{BC-c0zero}).}
 \label{Vphi-beta-c0zero}
\end{figure}

From the studies of heavy quark potential, we have found that a
negative $\sigma$ is favored. With the parameters $L=1 {\rm
GeV}^{-1}$, $\sigma_s=0.38 {\rm GeV}^{-2}$, $c_0=0$, and
$\sigma=-0.22 {\rm GeV}^2$, the yielded heavy quark potential is
close to the Cornell potential. In this subsection, we solve its
gravity dual and investigate the $\beta$ function of this model.

To solve the dilaton field from Eq.~(\ref{Ephiz}), we need to choose
two boundary conditions. For one of the boundary conditions, we use
the value of QCD running coupling at $3 {\rm GeV}$ as input, i.e,
$\alpha(E=3 {\rm GeV}) = 0.25$, which can be read from \cite{PDG},
and solve $z$ from Eq.(\ref{Energy_gauge_2}), this gives one
boundary condition $\phi(z=0.87)={\rm log}(0.25)$. It is noticed
that $3 {\rm GeV}$ is about the charmonium mass which is in the IR
region.

For another boundary condition, we can choose the same boundary
condition as in \cite{Galow:2009kw} by input the running coupling at
$8 {\rm GeV}$, which gives the boundary condition $\alpha(8
GeV)=0.18575$. However, because the produced $\beta$ function in
\cite{Galow:2009kw} is not a monotonic function, we guess this
strange behavior is induced by fixing two points of the running
coupling. Therefore, we choose to use the derivative of the dilaton
field at $z(E=3 {\rm GeV})=0.87$, i.e, $\phi'(z=0.87)$ as another
boundary condition. Because we don't know the value of the
$\phi'(z=0.8701)$, we choose it as a free parameter.

There are two types of boundary conditions we used:
\begin{eqnarray}
& & {\rm 1stBC}: \phi(z=0.87)={\rm log}(0.25), ~
\phi'(z=0.87)=0.9, \nonumber \\
& & {\rm 2ndBC}: \phi(z=0.87)={\rm log}(0.25), ~ \phi(z=0.38)={\rm
log}(0.18). \label{BC-c0zero}
\end{eqnarray}
where $\phi'(z=0.87)=0.9$ is used by the best fit of the QCD $\beta$
function, and $\phi(z=0.38)={\rm log}(0.18)$ is from the input of
running coupling $\alpha(8 GeV)=0.18$ at UV.

The dilaton field $\phi$ as a function of $z$ is shown in Fig.
\ref{phiz-c0zero}. It is found that for the two types of boundary
conditions, the solution of the dilaton field $\phi(z)$ is
monotonically increasing with $z$. For the 1st type of boundary
condition, $\phi$ increases more quickly with $z$ than the case with
2nd type boundary condition.

The dilaton potential $V_B(\phi)$ as a function of $\phi$ and the
$\beta$ function as a function of $\alpha$ are shown in Fig.
\ref{Vphi-beta-c0zero} (a) and (b), respectively. It is found that
for both types of boundary conditions, $V_B(\phi)$ decreases with
$\phi$, the dilaton potential in the IR regime is not bounded from
below, which might indicate a unstable vacuum. For the second type
boundary condition, i.e, the boundary condition used in
\cite{Galow:2009kw}, it is found that the produced $\beta$ function
is not a monotonic function of coupling $\alpha$. This behavior as
we have discussed, is due to the fixing running coupling constant at
two points. For the first type of boundary condition, the produced
$\beta$ function is monotonically decreasing with the coupling
constant $\alpha$, and it agrees reasonably well with the QCD
$\beta$ function, which is shown by dashed line in Fig.
\ref{Vphi-beta-c0zero}(b).

\subsubsection{The dilaton potential and the $\beta$ function for
positive $\sigma$}

\begin{figure}
\begin{center}
\epsfxsize=6.5 cm \epsfysize=6.5 cm \epsfbox{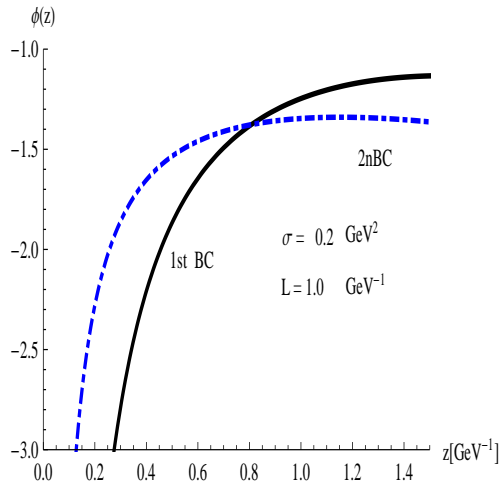}
\hspace*{0.1cm}
\end{center}
\caption[]{The dilaton field $\phi$ as a function of the the bulk
coordinate $z$ in the case of $L=1 {\rm GeV}^{-1}$, $\sigma_s=0.38
{\rm GeV}^{-2}$, $c_0=0$, and $\sigma=0.22 {\rm GeV}^2$. The
boundary conditions are described in
Eq.(\ref{BC-c0zero-sigmapositive}). } \label{Anti-phiz-c0zero}
\end{figure}

\begin{figure}
\begin{center}
\epsfxsize=6.5 cm \epsfysize=6.5 cm \epsfbox{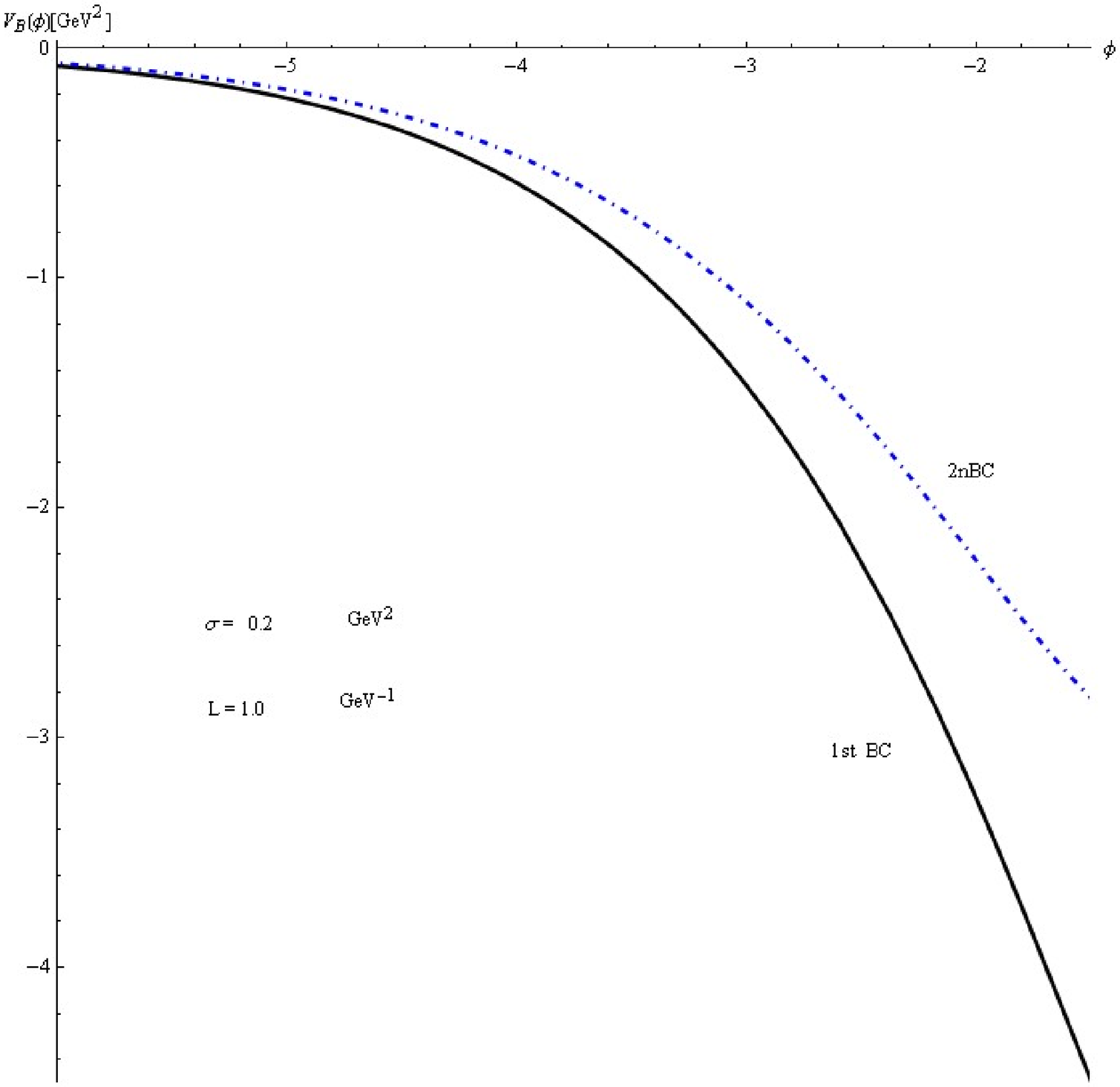}
\hspace*{0.1cm} \epsfxsize=6.5 cm \epsfysize=6.5 cm
\epsfbox{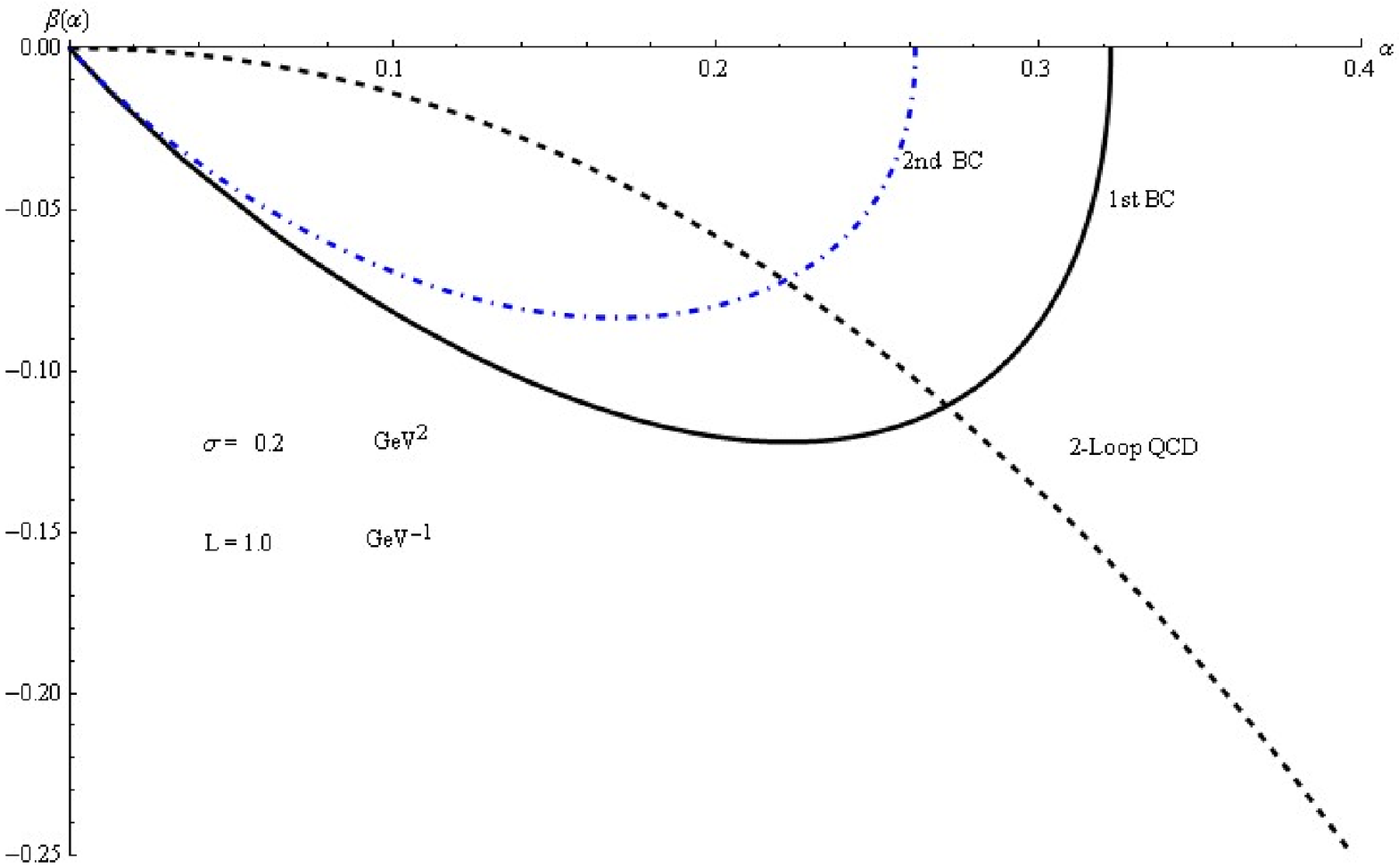} \vskip -0.05cm \hskip 0.15 cm
\textbf{( a ) } \hskip 6.5 cm \textbf{( b )} \\
\end{center}
\caption[]{ (a) The dilaton potential $V_B(\phi)$ as a function of
$\phi$ and (b) The $\beta$ function as a function of $\alpha$ in the
case of $L=1 {\rm GeV}^{-1}$, $\sigma_s=0.38 {\rm GeV}^{-2}$,
$c_0=0$, and $\sigma=0.2 {\rm GeV}^2$. The boundary conditions are
described in Eq.(\ref{BC-c0zero-sigmapositive}).}
 \label{Anti-Vphi-beta-c0zero}
\end{figure}

As a reference, we solve the gravity dual and investigate the
$\beta$ function for the case of positive $\sigma$, which
corresponds to the KKSS model or soft-wall model. The parameters for
the model are $L=1 {\rm GeV}^{-1}$, $\sigma_s=0.38 {\rm GeV}^{-2}$,
$c_0=0$, and $\sigma=0.2 {\rm GeV}^2$.

The two types of boundary conditions are
\begin{eqnarray}
& & {\rm 1stBC}: \phi(z=0.81)={\rm log}(0.25), ~
\phi'(z=0.81)=0.9, \nonumber \\
& & {\rm 2ndBC}: \phi(z=0.81)={\rm log}(0.25), ~ \phi(z=0.38)={\rm
log}(0.18),  \label{BC-c0zero-sigmapositive}
\end{eqnarray}
which are almost the same as Eq.(\ref{BC-c0zero}).

The dilaton field $\phi$ as a function of $z$ is shown in Fig.
\ref{Anti-phiz-c0zero}. It is found that for the two types of
boundary conditions, the solution of the dilaton field $\phi(z)$
monotonically increases to a maximum value at $z_m$. For the first
type boundary condition, $z_m=1.4 {\rm GeV}^{-1}$, and for the 2nd
type boundary condition, $z_m=1.0 {\rm GeV}^{-1}$. For both cases,
$z_m$ is much smaller than $z_{IR}$.

The dilaton potential $V_B(\phi)$ as a function of $\phi$ and the
$\beta$ function as a function of $\alpha$ are shown in Fig.
\ref{Anti-Vphi-beta-c0zero} (a) and (b), respectively. The dilaton
potential $V_B(\phi)$ decreases with $\phi$, which shows an unstable
potential. It is found that the $\beta$ function is very interesting
in the case of a positive $\sigma$. The $\beta$ function has two
fixed points where $\beta$ function vanishes: 1) one is a UV fixed
point, where $\beta=0$ when $\alpha=0$, 2) another is a IR fixed
point, where $\beta=0$ at a moderate strong coupling constant
$\alpha=0.26$ for 2ndBC and $\alpha=0.32$ for 1stBC, respectively.
If we take a lager value of $\sigma$, the IR fixed point will appear
at a smaller coupling constant, then we can have the Banks-Zaks
fixed point \cite{Banks-Zaks}.

\subsection{With both quadratic and logarithmic corrections}
\label{sect-c0=1}

\subsubsection{The heavy quark potential}

\begin{figure}[h]
\begin{center}
\epsfxsize=6.5 cm \epsfysize=6.5 cm \epsfbox{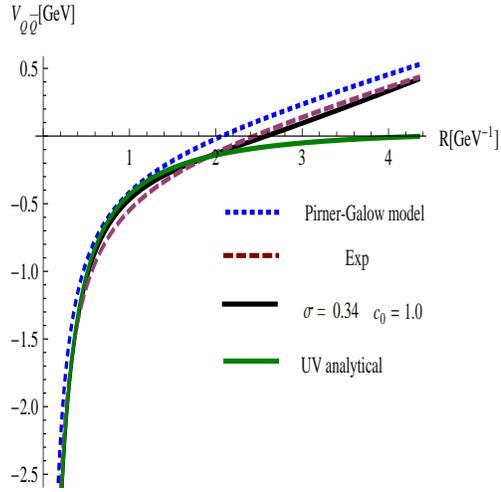}
\end{center}
\caption{ (a) The best fitted heavy quark potential as a function of
$R$ in the case of $c_0=1$ compared with Pirner-Galow result, the UV
analytical result and the Cornell potential.  In the case of
$c_0=1$, the other parameters are $L=1 {\rm GeV}^{-1}$,
$\sigma_s=0.38 {\rm GeV}^{-2}$, $\sigma=0.34 {\rm GeV}^2$ and
$z_{IR}=2.54 {\rm GeV}^{-1}$.  } \label{Vqq-Vphi-c0-one}
\end{figure}

From the Pirner-Galow metric Eq.(\ref{metric-PG}), we extract the
coefficient of $c_0=1$. Fig. \ref{Vqq-Vphi-c0-one} (a) shows the
heavy quark potential in the case of $c_0=1$, the best fitted result
(black solid line) gives $\sigma=0.34 {\rm GeV}^2$ and $z_{IR}=2.54
{\rm GeV}^{-1}$. With these parameters, the interquark distance and
the heavy quark potential diverges at $z_p=1.95 {\rm GeV}^{-1}$. It
is found that the heavy quark potential is perfectly fitted in the
regime $R<0.5 {\rm GeV}^{-1}$ and $R>2 {\rm GeV}^{-1}$, however, in
the regime $0.5 {\rm GeV}^{-1}<R<2 {\rm GeV}^{-1}$, the fitted heavy
quark potential is a little bit higher than the experimental data.

The result from the Pirner-Galow model is also shown in the figure
by using the short dashed line. It is found that the Coulomb part is
in good agreement with the Cornell potential, the linear part is
parallel to the Cornell potential, however, the value of $V_{Q{\bar
Q}}$ is a little bit higher than the data.

The green solid line is the UV analytical result from
Eq.(\ref{VQQ-UV}). It is found that this result is in good agreement
with the Coulomb part of the Cornell potential in the region $R<2
{\rm GeV}^{-1}$. The UV analytical is not valid any more above $2
{\rm GeV}^{-1}$.

\subsubsection{The dilaton potential and the $\beta$ function}
\label{Sec-Vphi-Beta-c01}

\begin{figure}[h]
\begin{center}
\epsfxsize=6.5 cm \epsfysize=6.5 cm \epsfbox{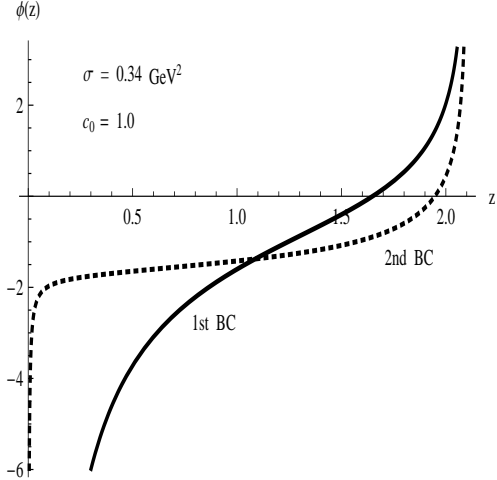}
\end{center}
\caption[]{The dilaton field $\phi$ as a function of the the bulk
coordinate $z$ in the case of $c_0=1$, $L=1 {\rm GeV}^{-1}$,
$\sigma_s=0.38 {\rm GeV}^{-2}$, $\sigma=0.34 {\rm GeV}^2$ and
$z_{IR}=2.54 {\rm GeV}^{-1}$. The two types of boundary conditions
are described in Eq.(\ref{BC-c01}). } \label{phiz-c0-one}
\end{figure}

\begin{figure}[h]
\begin{center}
\epsfxsize=6.5 cm \epsfysize=6.5 cm \epsfbox{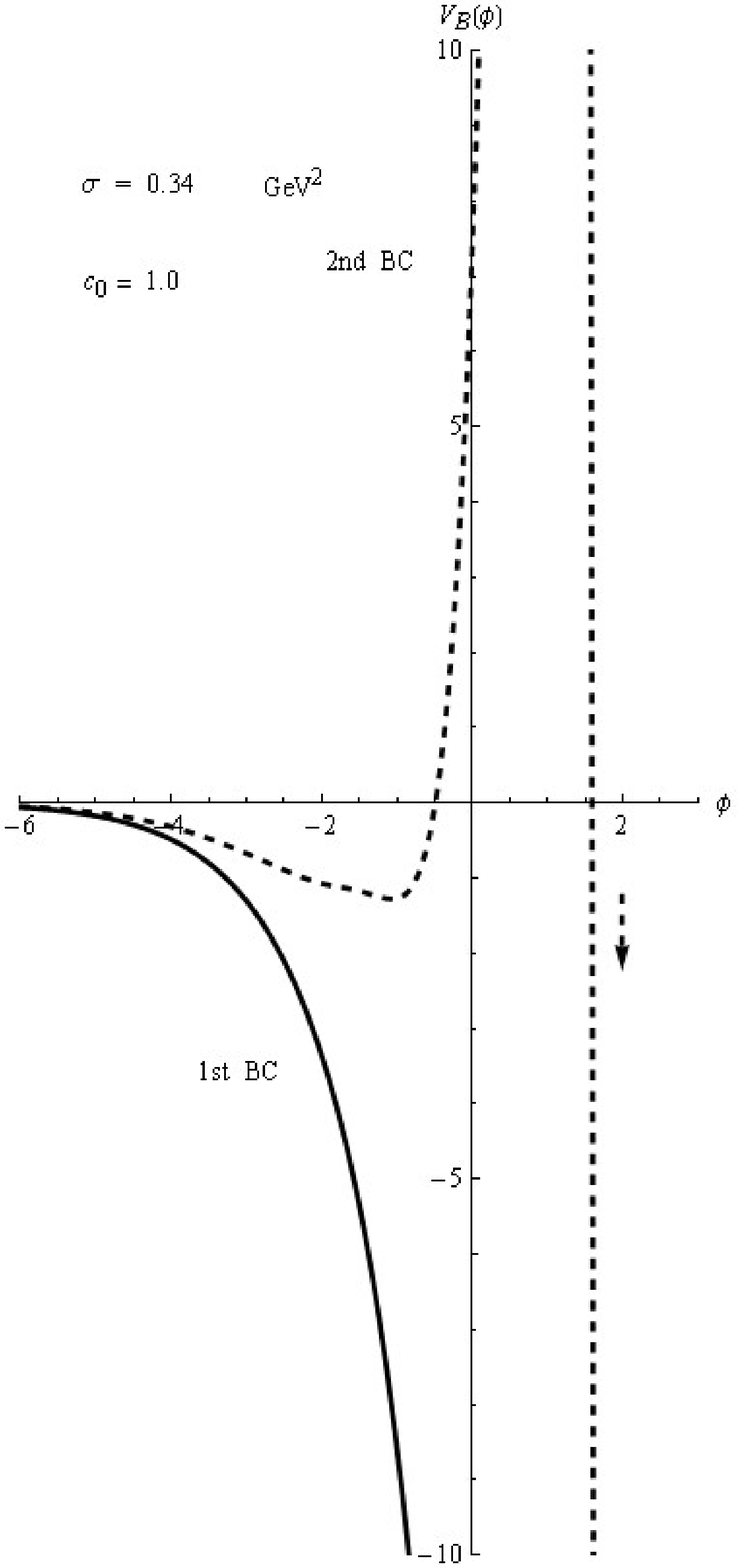}
\hspace*{0.1cm} \epsfxsize=6.5 cm \epsfysize=6.5 cm
\epsfbox{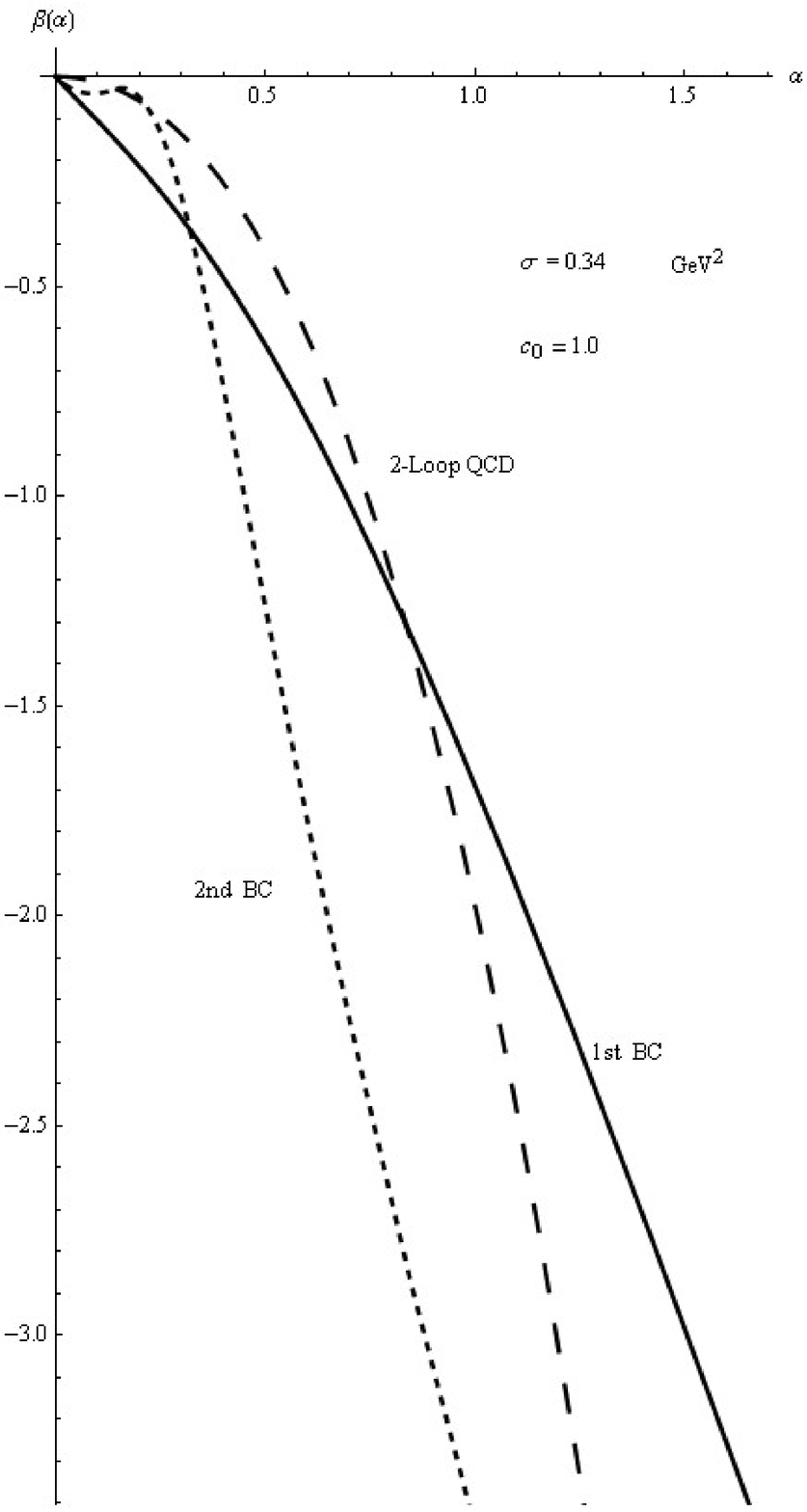} \vskip -0.05cm \hskip 0.15 cm
\textbf{( a ) } \hskip 6.5 cm \textbf{( b )} \\
\end{center}
\caption[]{(a) The dilaton potential $V_B(\phi)$ as a function of
$\phi$ and (b) the $\beta$ function as a function of coupling
constant $\alpha$ in the case of $c_0=1$, $L=1 {\rm GeV}^{-1}$,
$\sigma_s=0.38 {\rm GeV}^{-2}$, $\sigma=0.34 {\rm GeV}^2$ and
$z_{IR}=2.54 {\rm GeV}^{-1}$. The two types of boundary conditions
are described in Eq.(\ref{BC-c01}). } \label{Vphi-beta-c01}
\end{figure}

The dilaton field $\phi$ as a function of $z$ is shown in Fig.
\ref{phiz-c0-one} for two different type of boundary conditions:
\begin{eqnarray}
& & {\rm 1st BC}: \phi(z=1.08)={\rm log}(0.25), ~
\phi'(z=1.08)=2.5, \nonumber \\
& & {\rm 2nd BC}: \phi(z=1.08)={\rm log}(0.25), ~ \phi(z=0.42)={\rm
log}(0.18). \label{BC-c01}
\end{eqnarray}
Where $\phi(z=1.08)={\rm log}(0.25)$ is from the input of running
coupling $\alpha(E= 3{\rm GeV})=0.25$ at IR, $\phi'(z=1.08)=2.5$ is
by choosing the best fit of the $\beta$ function, and
$\phi(z=0.42)={\rm log}(0.18)$ is from the input of QCD running
coupling $\alpha(E=8{\rm GeV})=0.18$ at UV.

It is found that for these two types of boundary conditions, the
solution of the dilaton field $\phi(z)$ increases monotonically with
$z$. The difference lies in that $\phi$ is flat in a rather wide
region of $z$ for 2nd type of boundary condition.

The dilaton potential $V_B(\phi)$ as a function of $\phi$ and the
$\beta$ function as a function of $\alpha$ are shown in Fig.
\ref{Vphi-beta-c01} (a) and (b), respectively. It is found that for
the first type boundary condition, $V_B(\phi)$ is not bounded from
below in the IR region, however, the $\beta$ function monotonically
decreases with the increase of $\alpha$, which qualitatively agrees
with the behavior of QCD $\beta$ function. For the second type
boundary condition, it is found that the dilaton potential
$V_B(\phi)$ is unstable in the IR, however, the produced $\beta$
function is not a monotonic function of coupling $\alpha$. This
behavior as we have pointed out in Sec.\ref{betafunction}, is due to
the fixing of two points of running coupling constant.

\subsection{With only logarithmic correction}

\subsubsection{The heavy quark potential}
\label{sec-log-only}

\begin{figure}[h]
\begin{center}
\epsfxsize=6.5 cm \epsfysize=6.5 cm \epsfbox{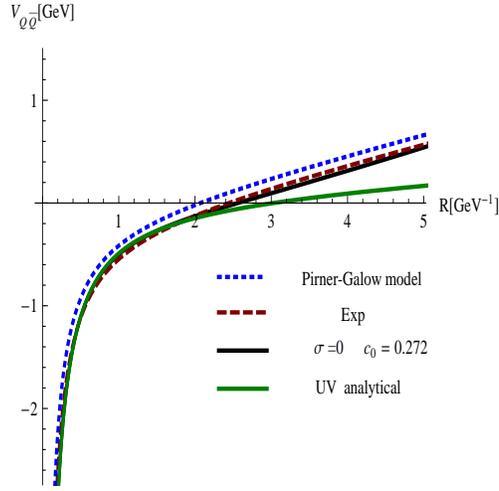}
\end{center}
\caption{ (a) The heavy quark potential as functions of the distance
$R$ in the case of $L=1 {\rm GeV}^{-1}$, $\sigma_s=0.38 {\rm
GeV}^{-2}$, $\sigma=0$ and $c_0=0.272$ and $z_{IR}=2.1 {\rm
GeV}^{-1}$. } \label{Vqq-Vphi-sigmazero}
\end{figure}

We now consider the case with only logarithmic correction when
$\sigma=0$. The best fitted heavy quark potential as functions of
quark anti-quark distance $R$ is shown in Fig.
\ref{Vqq-Vphi-sigmazero} (a) by using the black solid line.  The
results are compared with that from the Pirner-Galow model (short
dashed line) and the experimental data (the long dashed line) and
the UV analytical result. The best fit of the heavy quark potential
gives $c_0=0.272 {\rm GeV}^2$ and $z_{IR}=2.1 {\rm GeV}^{-1}$. With
these parameters, numerical calculations shows that the interquark
distance $R$ becomes divergent at $z_p=1.85 {\rm GeV}^{-1}$ (rough
estimates gives $z_{p}\sim \frac{2 z_{\text{IR}}}{c_0+2}$). It is
found that the heavy quark potential can be perfectly fitted in the
whole regime of $R$ with only logarithmic correction.

\subsubsection{The dilaton potential and the $\beta$ function}
\label{sec-Vphi-beta-sigmazero}

\begin{figure}[h]
\begin{center}
\epsfxsize=6.5 cm \epsfysize=6.5 cm \epsfbox{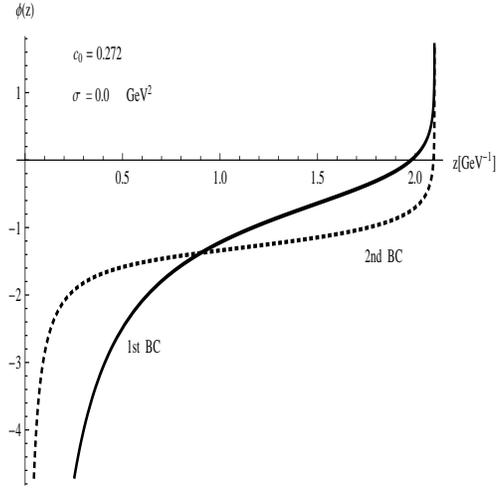}
\end{center}
\caption[]{The dilaton filed $\phi$ as a function of $z$ in the case
of $L=1 {\rm GeV}^{-1}$, $\sigma_s=0.38 {\rm GeV}^{-2}$, $\sigma=0$
and $c_0=0.272$ and $z_{IR}=2.1 {\rm GeV}^{-1}$. The boundary
conditions are described in Eq.(\ref{BC-sigmazero}).}
\label{phiz-sigmazero}
\end{figure}

\begin{figure}[h]
\begin{center}
\epsfxsize=6.5 cm \epsfysize=6.5 cm \epsfbox{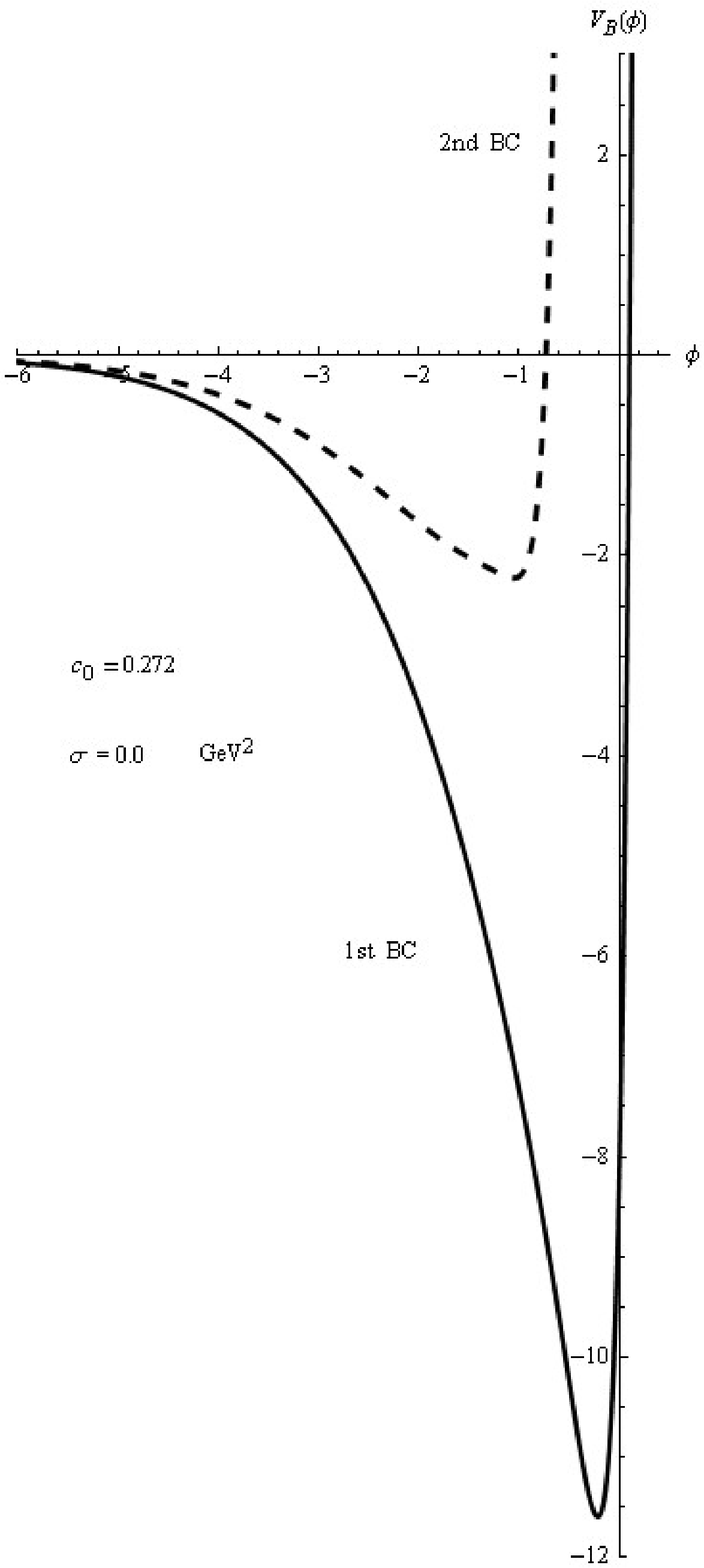}
\hspace*{0.1cm} \epsfxsize=6.5 cm \epsfysize=6.5 cm
\epsfbox{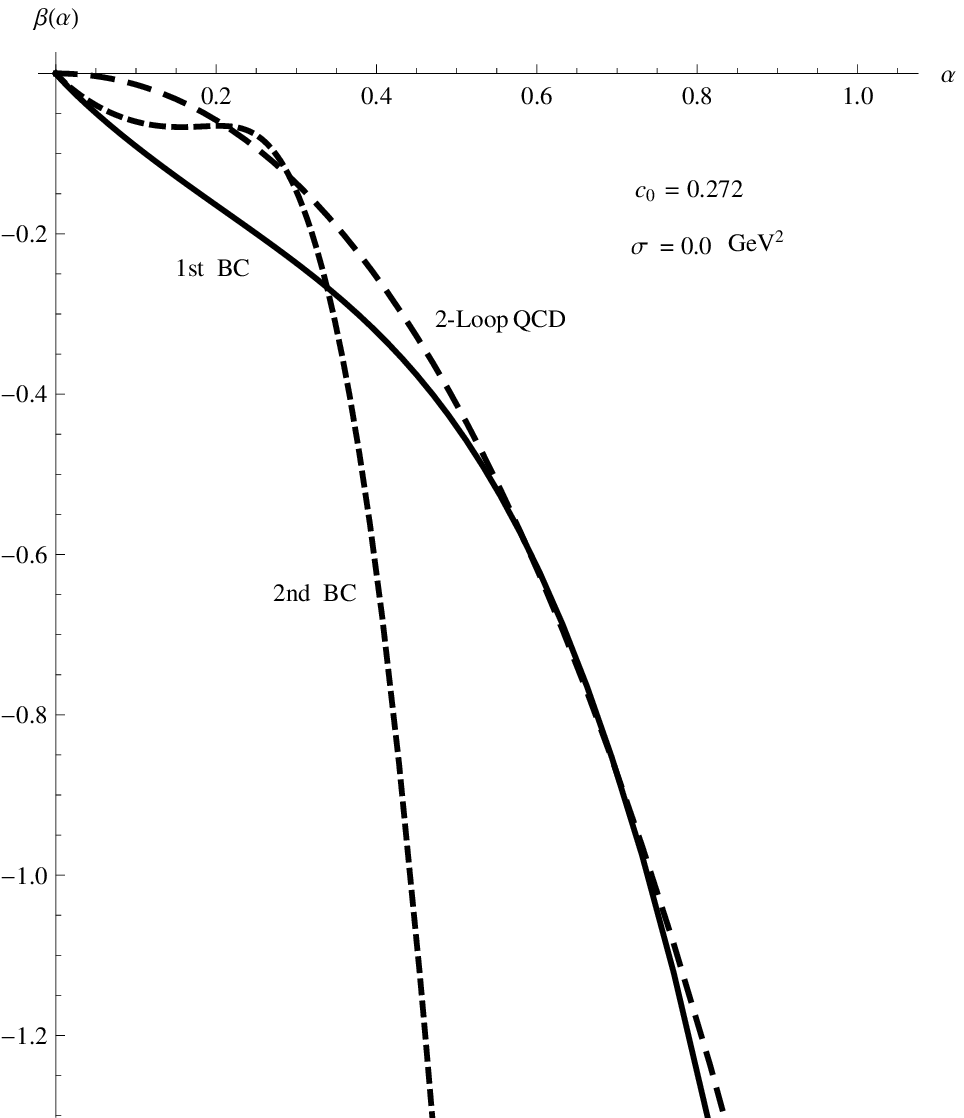} \vskip -0.05cm \hskip 0.15 cm
\textbf{( a ) } \hskip 6.5 cm \textbf{( b )} \\
\end{center}
\caption[]{(a) The dilaton potential $V_B$ as a function of $\phi$
and (b) the $\beta$ function as a function of coupling constant
$\alpha$ in the case of $L=1 {\rm GeV}^{-1}$, $\sigma_s=0.38 {\rm
GeV}^{-2}$, $\sigma=0$ and $c_0=0.272$ and $z_{IR}=2.1 {\rm
GeV}^{-1}$. The boundary conditions are described in
Eq.(\ref{BC-sigmazero}).}
 \label{Vphi-beta-sigmazero}
\end{figure}

The dilaton field $\phi$ as a function of $z$ is shown in Fig.
\ref{phiz-sigmazero} for two different type of boundary conditions:
\begin{eqnarray}
& & {\rm 1st BC}:  \phi(z=0.9)={\rm log}(0.25), ~
\phi'(z=0.9)=1.7, \nonumber \\
& & {\rm 2nd BC}:  \phi(z=0.9)={\rm log}(0.25), ~ \phi(z=0.39)={\rm
log}(0.185). \label{BC-sigmazero}
\end{eqnarray}
Where $\phi(z=0.9)={\rm log}(0.25)$ is from the input of the
$\alpha(E=3 {\rm GeV})=0.25$ at IR, $\phi'(z=0.9)=1.7$ is determined
by choosing the best fitting of QCD $\beta$ function, and
$\phi(z=0.39)={\rm log}(0.18)$ is from $\alpha(E=8{\rm GeV})=0.18 $
at UV.

It is found that for these two types of boundary conditions, the
solution of the dilaton field $\phi(z)$ increases monotonically with
$z$. The difference lies in that $\phi$ is flat in a rather wide
region of $z$ for 2nd type of boundary condition.

The dilaton potential $V_B(\phi)$ as a function of $\phi$ and the
$\beta$ function as a function of $\alpha$ are shown in Fig.
\ref{Vphi-beta-sigmazero} (a) and (b), respectively. It is found
that for the second type boundary condition, the dilaton potential
$V_B(\phi)$ is stable which is bounded from below in the IR,
however, the produced $\beta$ function is not a monotonic function
of coupling $\alpha$. This behavior as we have discussed, is due to
the fixing of two points of running coupling constant. For the first
type boundary condition, $V_B(\phi)$ is also a stable potential
which is more deeply bounded from below in the IR region, moreover,
the $\beta$ function monotonically decreases with the increase of
$\alpha$, and coincides with the QCD $\beta$ function in IR regime.

It is noticed that in the case with only logarithmic correction, the
5D dilaton potential $V_B(\phi)$ has the same shape as the dilaton
potential in an effective 4D QCD model in \cite{shifman-report}.

\section{The more compact model with only logarithmic corrections}

From studies in previous section, it is found that the model with
only logarithmic correction in the deformed warp factor can fit the
heavy quark potential perfectly, which is much better than the model
with only quadratic correction. It might not be a surprise because
there are four parameters used, i.e, the deformed ${\rm AdS}_5$
radius $L$, the string tension $\alpha$, the coefficient $c_0$ and
the IR cut-off $z_{IR}$, while for the model with only quadratic
correction, there are only three parameters, i.e, the deformed ${\rm
AdS}_5$ radius $L$, the string tension $\sigma_s$, and the
coefficient $\sigma$.

It should be mentioned that for the case of $c_0=1$ in Sec.
\ref{sect-c0=1}, five parameters have been used to fit the heavy
quark potential, and the best fitted result is better than the model
with only quadratic correction, but not as good as the model with
only logarithmic correction. Remind of the results in Ref.
\cite{White:2007tu}, White found that the model with only quadratic
correction, which has less parameters are better than the
backreaction model, which has more parameters to produce the heavy
quark potential. Therefore, it is not necessarily correct that one
can fit the three parameters in the Cornell potential with enough
parameters.

Still, we hope to improve our model with only logarithmic
correction. It is found there are two length scales in the model,
i.e, the deformed ${\rm AdS}_5$ radius $L$ and the IR cut-off
$z_{IR}$. We can combine these two length scales into one, and
choose the following metric structure:
 \bea
ds^2=G_{\mu\nu}dX^\mu dX^\nu&=&e^{2\mathcal {A}(z)}\left(
dt^2+d\vec{x}^2+dz^2\right)\nonumber\\
&=&\frac{h(z)z_{IR}^2}{z^2}\left(
dt^2+d\vec{x}^2+dz^2\right)\label{metircmodel}
 \eea
with ${A}(z)$ and $h(z)$ taking the following expressions:
 \bea \mathcal
{A}(z)=-\log(\frac{z}{z_{IR}})-\frac{c_0}{2}\log(\frac{z_{IR}-z}{z_{IR}})
 \eea
\bea h(z)=\exp\left( -c_0\log(\frac{z_{IR}-z}{z_{IR}})\right) \eea

Following the same procedure, we obtain the expression of the
renormalized heavy quark potenital $V_{Q\bar{Q}}^{ren.}$ in the form
of
\begin{eqnarray}\label{Vregular2}
\begin{split}
V_{Q\bar Q}^{ren.}(z)&=-\frac{1}{\pi\sigma_s}\frac{z_{IR}^2}{z}
+\frac{1}{\pi\sigma_s}\frac{z_{IR}^2}{z}\int_0^1 d\nu \left
(\frac{h(\nu z)}{\nu^2}\vphantom{\frac{1}{\sqrt{1-\nu^4 \left(
\frac{h(z)}{h(\nu z)} \right)^2}}-\frac{1}{\nu^2}}\right. \left.
\frac{1}{\sqrt{1-\nu^4\left ( \frac{h(z)}{h(\nu z)}\right
)^2}}-\frac{1}{\nu^2}- \frac{c_0 z}{z_{IR}\nu}\right ),\end{split}
\end{eqnarray}
 and the interquark distance $R$ has the form of
\bea R(z) &=&2 z \int_0^1 d\nu \frac{e^{2\mathcal
{A}(z)}}{e^{2\mathcal {A}(\nu z)}}\frac{1}{\sqrt{1- \left
(\frac{e^{2\mathcal {A}(z)}}{e^{2\mathcal {A}(\nu z)}}\right )^2}}.
\label{distance1} \eea

The UV limit of the heavy quark potential has expression of
\begin{eqnarray}
V_{Q\bar Q}(R)&=&-\frac{0.23 z_{\text{IR}}^2}{\sigma_s R}+\frac{0.17
c_0 z_{\text{IR}}}{\sigma_s }+\frac{\left(0.22 c_0+0.24 c_0^2\right)
R}{\sigma_s }. \label{UV-VQQ-compact}
\end{eqnarray}

The Coulomb part can be fitted very well with the string tension
$\sigma_s=1.6 {\rm GeV}^{-2}$, which is just $z_{IR}^2/L^2$ times
$\sigma_s=0.38$ in Sec.\ref{sec-log-only}. The best fit of the heavy
quark potential gives $c_0=0.272 {\rm GeV}^2$ and $z_{IR}=2.11 {\rm
GeV}^{-1}$, which are the same as those in Sec.\ref{sec-log-only}.
The results of $\phi(z)$, $V_B(\phi)$ and $\beta(\alpha)$ in the
compact model are almost the same as those in Sec.
\ref{sec-Vphi-beta-sigmazero}, therefore, we neglect the figures in
this part.

The advantage of the compact model is that with only three
parameters, we can produce the results of heavy quark potential and
QCD $\beta$ function as good as those in the model with four
parameters.

\section{Conclusion and discussion}

In this paper, we study a \textit{holographic} QCD model which
contains a quadratic term $ -\sigma z^2$ and a logarithmic term
$-c_0\log[(z_{IR}-z)/z_{IR}]$ with an explicit infrared cut-off
$z_{IR}$ in the deformed ${\rm AdS}_5$ warp factor. We investigate
the heavy quark potential, solve the dual gravity with dilaton field
in G{\"u}rsoy -Kiritsis-Nitti (GKN) framework, and study the
corresponding $\beta$ function for three cases, i.e, with only
quadratic correction, with both quadratic and logarithmic
corrections and with only logarithmic correction. Our studies show
that in the case with only quadratic correction, the heavy quark
potential can be qualitatively fitted with a negative $\sigma$, and
the beta-function agrees with the QCD beta-function reasonably well,
however, the dilaton potential is unbounded in infrared regime. In
the case with only logarithmic correction, the heavy quark Cornell
potential can be fitted very well, the corresponding beta-function
agrees with the QCD beta-function at 2-loop level reasonably well,
and the dilaton potential is bounded from below in infrared. We also
propose a more compact model which has only logarithmic correction
in the deformed warp factor, which can describe the heavy quark
potential and QCD $\beta$ function very well with only three
parameters.

{\bf Stability analysis of the dilaton potential $V_B$}

From our numerical studies, it is shown that the dilaton potential
$V_B(\phi)$ for the case with only quadratic correction keeps
decreasing with $\phi$, which indicates the potential is unstable.
For the case with only logarithmic correction, the dilaton potential
firstly decreases with $\phi$ then moves upward in the IR regime,
which indicates that the dilaton potential is stable. In the
following, we analyze the stability of the dilaton potential for the
given metric Eq.(\ref{metric-firstmodel}).

Because $\phi$ monotonically increases with $z$, we can analyze the
stability of $V_B(\phi)$ from $V_B(z)$ in Eq.(\ref{DilatonVphi}).
Substitute Eq.(\ref{Ephiz}) into Eq.(\ref{DilatonVphi}), we can have
the expression as:
\begin{equation}
V_B(z)=-e^{-2A(z)} \left(12(A'(z))^2-\frac{4}{3}(\phi'(z))^2
\right). \label{DilatonVphi-2}
\end{equation}
For the metric structure Eq.(\ref{metric-firstmodel}) in the string
frame with quadratic and logarithmic correction, the metric in the
Einstein frame takes the explicit form of
\begin{eqnarray}
A(z)=-\frac{2}{3}\phi - \mathcal{A}(z), ~  \mathcal{A}(z)={\rm log}z
+\frac{1}{4}\sigma z^2+\frac{c_0}{2}{\rm
log}\frac{z_{IR}-z}{z_{IR}},
\end{eqnarray}
and the dilaton potential Eq.(\ref{DilatonVphi-2}) becomes
\begin{equation}
V_B(z)=-4
e^{\frac{4}{3}\phi+2\mathcal{A}}[(\phi')^2+4\phi'\mathcal{A}'+3(\mathcal{A}')^2
]. \label{DilatonVphi-3}
\end{equation}
Because the two square terms in the bracket are always positive,
also from our numerical results for the physical cases, we have
$\phi'>0$, the only chance for $V_B(z)$ to change sign in the IR
regime is to have a negative derivative of $\mathcal{A}$. From the
explicit expression of $\mathcal{A}'$, i.e,
\begin{equation}
\mathcal{A}'=\frac{1}{z}+\frac{\sigma z}{2}-\frac{c_0}{2(z_{IR}-z)},
\end{equation}
we can read that with only quadratic correction, i.e, when $c_0=0$,
if $\sigma>0$, $\mathcal{A}'$ is always positive. When $\sigma<0$,
e.g, $\sigma=-0.22$ to produce the Cornell potential, $\mathcal{A}'$
is also positive in the regime of $z<z_{IR}$. In the case with only
logarithmic correction, i.e, when $\sigma=0$, we can see that in the
IR regime when $z \rightarrow z_{IR}$, $\mathcal{A}'\rightarrow
-\infty$ when $c_0>0$, therefore, $V_B(z)$ might change sign and
become positive in the IR regime.

{\bf Positive or negative quadratic correction?} To fit the heavy
quark potential and to produce the QCD $\beta$ function, a negative
$\sigma$, i.e, the Andreev model is favored. However, from our
previous experience in Ref.\cite{Dp-Dq}, to produce the Regge
behavior of $\rho$ meson, a positive $\sigma$ is needed, i.e, the
KKSS model or soft-wall model is favored. One possible explanation
is that different model is needed to describe the physics in light
flavor sector and heavy flavor sector, respectively.  The subtlety
of the quadratic correction in the \textit{holographic} model
deserves further careful studies in the future. In our future
project, we will check whether Regge behavior can be described in
the holographic model with only logarithmic correction.

One interesting result in the KKSS model is that the corresponding
$\beta$ function exists a IR fixed point. If the fixed point exists
at a strong coupling regime, the KKSS model might be interesting to
study the unitary regime of BCS-BEC crossover in cold atom system.
If the fixed point exists at a weak coupling regime, which might be
interesting to study the dynamical electroweak symmetry breaking
physics.

\vskip 1cm \noindent {\bf Acknowledgments}:

The authors thank B.Galow, P. Hohler, F. Jugeau, M.S. Ma, S. Pu, M.
Stephanov, N. Su, F.K. Xu, Y. Yang, and H.Q. Zhang for valuable
discussions. The work of M.H. is supported by CAS program
"Outstanding young scientists abroad brought-in", CAS key project
KJCX3-SYW-N2, NSFC10735040, NSFC10875134, and K.C.Wong Education
Foundation, Hong Kong.

\end{document}